\documentclass[aps,prl,reprint,superscriptaddress,amsmath,amssymb]{revtex4-1}

\usepackage{graphicx}
\usepackage{dcolumn}
\usepackage{booktabs}
\usepackage{amssymb}
\usepackage{amsthm,amsmath}
\usepackage{bm}

\usepackage{tabularx}

\begin{document}

\preprint{APS/123-QED}


\title{Optimizing Laser Wakefield Acceleration in the Nonlinear\\ Self-Guided Regime for Fixed Laser Energy}


\author{A. Davidson}
\email{davidsoa@ucla.edu; Current address at: U.S. Naval Research Lab, 4555 Overlook Ave. SW, Washington, DC
20375.}
\author{A. Tableman}
\author{P. Yu}
\author{W. An}
\author{F. S. Tsung}
\affiliation{University of California, Los Angeles, CA 90095, USA}
\author{W. Lu}
\affiliation{Tsinghua University, Beijing, China}
\author{R. A. Fonseca}
\affiliation{GoLP/Instituto de Plasmas e Fus\~{a}o Nuclear, Instituto Superior T\'ecnico, Universidade de Lisboa, Lisbon, Portugal}
\affiliation{DCTI/ISCTE - Instituto Universit\'{a}rio de Lisboa, Lisbon, Portugal}
\author{W.B. Mori}
\affiliation{University of California, Los Angeles, CA 90095, USA}


\date{\today}

\begin{abstract}
The scaling laws for laser wakefield acceleration in the nonlinear, self-guided regime [Lu et al. Phys. Rev. Spec. Top.  Accel. Beams 10, 061301 (2007)] are examined in detail using the  quasi-3D version of the particle-in-cell code OSIRIS. We find that the scaling laws continue to work well as the plasma density is reduced while the normalized laser amplitude is kept fixed. For fixed laser energy, the energy gain of an isolated bunch of electrons can be improved with some loss in the bunch charge by shortening the normalized pulse length until self-guiding no longer occurs, and through the use of asymmetric longitudinal profiles with rapid rise times. For example,  without any external guiding a $15$ J,$.8\mu m$ laser  with a pulse length of
$46 fs$ ($39 fs$)
is found to generate a quasi-mono-energetic bunch of 355pC (227pC) with a max energy of  $3.25$ ($4.04$) GeV with an acceleration distance of 2.43 cm (3.08cm). Furthermore, a bunch with  39.4pC and a maximum energy  $4.6 GeV$ is 
produced for an asymmetric laser with a rapid rise time. Studies for $30$ J and $100$ J lasers are also presented.
The implications of these results for more controlled injection methods is also discussed. 
\end{abstract}

\pacs{insert PACS here}

\maketitle

\section{Introduction}

Plasma-based acceleration (PBA) \cite{Litos14,Leemans06, Blumenfeld07, Kneip09, Froula09, Clayton10, Wang13, Kim13, Leemans14}
 has received much recent attention owing to its potential to lead to a new generation of compact accelerators that could lead to a smaller and lower cost linear collider and coherent x-ray source. In PBA either an intense laser or particle beam drives a plasma wave wakefield before it pump depletes as it traverses a tenuous plasma. Electrons or positrons are then loaded into these wakefields and are then accelerated with gradients in excess of 10 GeV/m. When the wakefield is driven by a  laser or a particle beam the process is called Laser Wakefield Acceleration (LWFA) or Plasma Wakefield Acceleration (PWFA) respectively\cite{Esarey96} . 

In PWFA and LWFA the wakefields can be excited in linear or nonlinear regimes. To date, many of the LWFA and PWFA experiments that have demonstrated electron energies exceeding 100 MeV \cite{Litos14,Leemans06, Blumenfeld07, Kneip09, Froula09, Clayton10, Wang13, Kim13, Leemans14} have operated in the nonlinear multi-dimensional wakefield regime. In this regime, which is sometimes referred to as the bubble or blowout regime, the wake is excited by the laser or particle beam expelling essentially all of the plasma electrons sideways where they then flow backwards in a narrow sheath which surrounds an ion cavity (the more massive neutralizing ions do not move). The ions then pull the electrons in the sheath back towards the axis thereby creating a wakefield. The fields inside this wakefield are electromagnetic in character and can be completely described by a the gauge invariant wake potential $\psi=(\phi-A_z)$ (cgs units) where $\phi $ is the scalar potential and $A_z$ is the component of the vector potential in direction that the wake is moving\cite[and references therein]{weilu2, Mora97,Esarey96}. In this case $\psi$ depends on the variable $\xi=(ct-z)$. The accelerating field ($E_z$) and focusing field ($(\vec E + \hat z x \vec B)_{\perp})$) on a particle moving near c in the $\hat z$ direction are given  by $\nabla_{\xi} \psi$ and $\vec \nabla_{\perp} \psi$ respectively. These fields have ideal properties for accelerating electrons, i.e., the accelerating field does not depend on $x_{\perp}$, the focusing field points in the radial direction and depends linearly on $x_{\perp}$ and it does not depend on $\xi$. As a result nonlinear wakes are ideal candidates for acceleration electrons in a linear collider and for generating GeV class beams for use in a next generation XFEL.  

The acceleration gain within a single PBA stage will scale with the acceleration gradient times the acceleration length. In the nonlinear blowout regime the acceleration gradient is well understood. The acceleration length is the smaller between the pump depletion length, the diffraction length, or the dephasing length. In 2007 Lu et al. \cite{weilu} presented a phenomenological description of LWFA in the nonlinear regime where the bulk of the laser is self-guided by the electron sheath. A significant amount of the leading edge of the laser pump locally pump depletes as it creates the wake before it diffracts. As it pump depletes the edge of the laser erodes backwards (etching speed) which leads to phase velocity of the wake less than the linear group velocity. Using a mixture of theory and simulations, parameter dependencies of these phenomena were developed and then combined into scaling laws for the energy gain in terms of the laser power, plasma density, and laser wave length, giving
\begin{align}
\label{elestimate}
\Delta E~[GeV] \simeq \biggl(\frac{1.7 \cdot P}{100TW}\cdot \frac{0.8}{\lambda_o[\mu m]}\biggr )^{1/3}\biggl(\frac{10^{18}}{n_{p} [cm^{-3}]}\biggr)^{2/3}.
\end{align}
Importantly, the scaling laws implicitly assume the laser spot size is matched to the maximum blowout radius, $w_0=2 a_0^{1/2} c/\omega_p$ and the pulse length is matched to the etching distance,  $c \tau=2/3 w_0$.  Here, $a_0$ is the normlized vector potential of the laser, $eA/mc^2$. As described in ref. \cite{weilu}, this regime is distinct from the work of Pukhov and Meyer-ter-vehn \cite{Pukhov02}and Gordienko and Pukhov \cite{Gordienko05} which is often called the bubble regime, where much higher laser intensities and plasma densities were considered. In the bubble regime, self-trapping of copious electrons cannot be avoided and the resulting loaded wakefield is not attractive for generating high quality electron beams.

A laser can be guided with a stable spot size by a plasma channel with a parabolic density profile with its minimum on the laser's axis\cite{Mori97}. Even without an external channel, however, the relativistic mass corrections may potentially give the same effect if the power is above some critical power\cite{Sun87}. At the front of the laser, however, the density compression from the ponderomotive force of the laser cancels out the increase in the index of refraction from relativistic mass effects, leading some to believe that a short pulse laser with $\tau \lesssim 1/\omega_p$ cannot be self-guided\cite{Esarey96}. On the contrary, Ref.~\cite{Decker96} shows that for a high enough power self-focusing may be possible because the leading edge of the laser is continuously locally pump depleted before it diffracts, while the back of the pulse, located behind the density compression, is still guided.

The simulations presented in ref.~\cite{weilu}  showed that a properly matched laser pulse remained self-guided for up to $\approx 5$ $Z_R$ where  $Z_R=\pi W_0^2/\lambda$ is a Rayleigh length. Self-guiding was also demonstrated in experiments\cite{SelfGuideExp}. However, there remains questions whether self-guiding will continue to scale as the plasma density is lowered and the acceleration length increases in units of $Z_R$. 

In this paper, we show that LWFA in the nonlinear self-guided regime can indeed be scaled to much higher energies. We use a new quasi-3D algorithm\cite{Lifschitz09,davidson15,Yu15} in the particle-in-cell code OSIRIS\cite{OSIRIS02} to carry out an extensive parameter scan at lower densities and higher laser energies than were originally studied. We confirm that self-guiding still occurs. We also recast the scaling laws in terms of laser energy rather than laser power and show that the electron energy can be optimized by shortening the laser pulse and changing is longitudinal profile. These new results predict that using present day 15 to 30 Joule lasers it is possible to generate 5 to 8 GeV electrons respectively without the need for any external guiding.  For simplicity, we have considered situations where the accelerated electrons are self-trapped in uniform plasmas through the evolution of the size of the wake\cite{Esarey96}.

\begin{figure}[t] 
   \centering
   \begin{tabular}{l}
      \includegraphics[width=8.0cm]{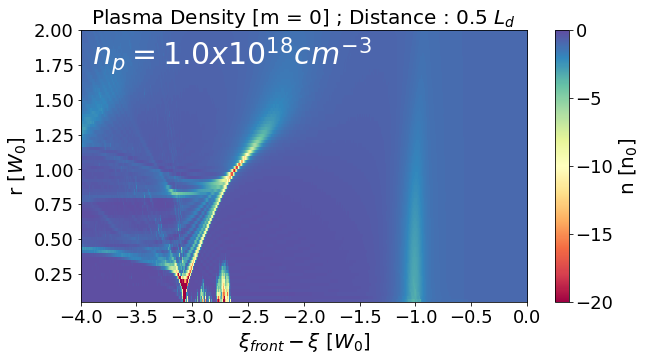} \\
      \includegraphics[width=8.0cm]{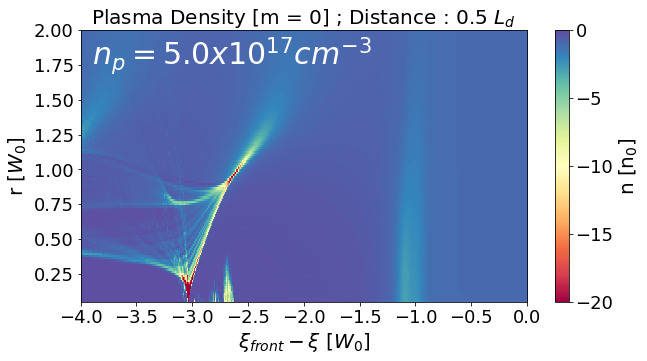} \\ 
      \includegraphics[width=8.0cm]{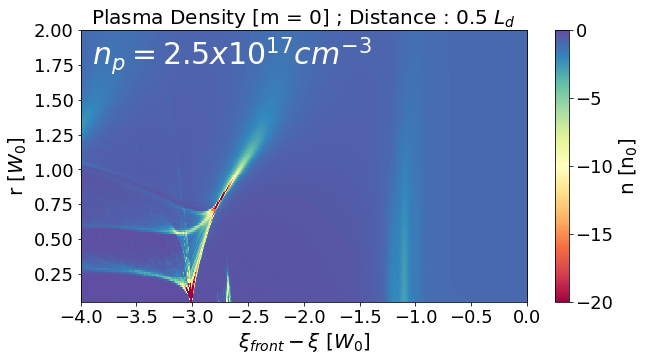}
   \end{tabular}
   \caption{Here we compare the density profile of three scaled LWFA simulations in normalized units. The density (color) axis is normalized to the initial plasma density of each case, and the $r$ and $\xi$ coordinates are plotted in units of the scaled initial spot size, $W_0$. The features of the bubble scale qualitatively very well as we scale to lower plasma densities and higher particle energies. }
   \label{densityscaling}
\end{figure}

\section{Simulation Approach}

Performing LWFA simulations in the nonlinear blowout regime in a full 3D simulation for electron energies beyond a few GeV energies quickly become computationally expensive and eventually unfeasible as it scales as the square of the output electron energy . However, we have recently implemented and improved \cite{davidson15} a hybrid PIC code which is PIC in (r,z) and gridless in the azimuthal mode number, m. For a linearly polarized laser with a nearly symmetric spot size  only the m=0 and m=1 azimuthal modes need to be kept \cite{Lifschitz09}. This reduces the computational needs by a factor of roughly the number of grids in the transverse direction. In ref \cite{davidson15} it is also shown that this quasi-3D algorithm can provide  quantitative agreement to results from the full 3D PIC algorithm. 


\begin{table}[htbp]
   \centering
   \begin{tabular}{lccccccc} 
   \hline
      \multicolumn{2}{c}{Est. Particle Steps}  & Power &  $n_p$ & $W_0$ & $L_d$ & $a_0$ & $\Delta E$\\
      -3D-          &          -Quasi-3D- & (TW)   &  ($cm^{-3}$) & ($\mu m$) & ($cm$) & & (GeV) \\ 
      \hline
      $1.9e15$  &  $8.5e12$  & 200 & 1.5e18 & 19.5 & 1.5\footnote{Lu et al. conducted this simulation over $0.75$ cm, and not the entire dephasing length, $L_d$.}  & 4.0 & 1.58\\
      $6.0e15$   &  $2.1e13$  & 324     & 1.0e18 & 22.0 & 2.62 & 4.44 & 2.52 \\
      $4.6e16$  & $1.2e14$ & 649  & 5.0e17 & 31.7 & 7.37 & 4.44 & 5.28 \\
      $3.6e17$  &  $6.6e14$   & 1298 & 2.5e17 & 44.8 & 20.8 & 4.44 & 10.57 \\
      \vdots & \vdots   & \vdots &  \vdots & \vdots & \vdots & \vdots & \vdots\\
      \hline
   \end{tabular}
   \caption{A list of laser and plasma parameters for simulations that could test the scaling laws of \cite{weilu}. The table includes estimates of the number of particle steps required for full 3D or quasi-3D OSIRIS simulations.}
   \label{cpuhourstable}
\end{table}

To better illustrate the advantage of using a quasi-3D geometry for our PIC simulations, estimates for the number of particle steps for a sequence of such simulations are presented in Table~\ref{cpuhourstable}. The required number of CPU hours would be proportional to the number of particle steps, if parallel scalability and load balancing are ideal (in effect they provide a lower estimate on the necessary CPU hours). The estimates were calculated by assuming a ``standard" resolution of $\Delta z = 0.2 k_0^{-1}$, $\Delta r = 0.1 k_p^{-1}$, and a box size of about $5.2 W_0$ in the longitudinal direction (width comparable to the original Lu et al. runs), and a transverse box size equal to the initial spot size times the total number of Rayleigh lengths over the dephasing lengths ($L_d W_0 / Z_R$). We also assumed  four particles were initialized per cell in the 3D Cartesian case. If the simulation is operated on a machine that takes an average of $500 ns$ for a typical LWFA simulation, the third and fourth rows of Table~\ref{cpuhourstable} correspond to 3 million and 26 million cpu hours, respectively. The number of particle steps scales as the maximum electron energy gain cubed. Therefore, it is not currently feasible to examine the accuracy of the scaling laws for lower plasma density with a full 3D Cartesian PIC code. 

2D cylindrical geometry simulations are often pursued for beam-driven PWFA problems, but they are not effective for laser-driven problems due to the fact that for typical linear or circularly polarized lasers the laser field themselves are not cylindrically symmetric. However, these laser fields are captured by the $m=1$ mode present in the quasi-3D description. The quasi-3D algorithm allows us to push scaling laws to previously unexplored regimes while retaining the important three dimensional physics.

\begin{figure}[htbp] 
   \centering
   \begin{tabular}{c}
 \includegraphics[width=9.0cm]{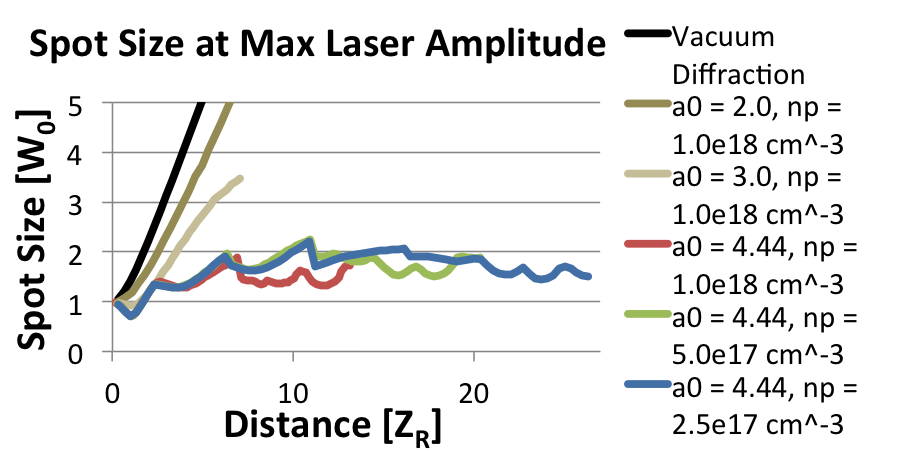} \\
 \includegraphics[width=8.0cm]{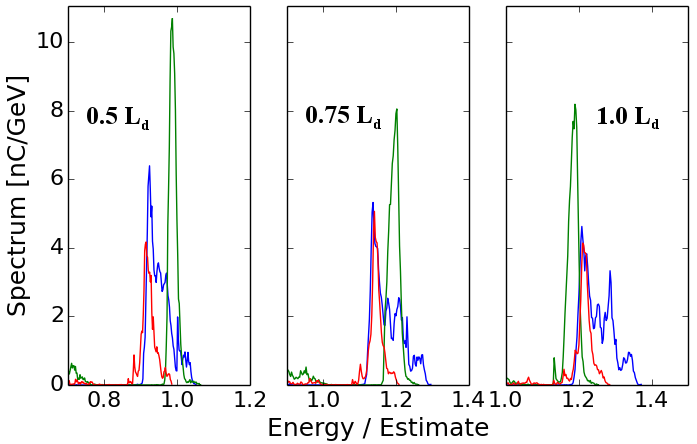}
   \end{tabular}
   \caption{(top) The evolution of the spot sizes at the location of the maximum laser amplitude is plotted over distance in Rayleigh lengths. Results for vacuum diffraction and simulations with $a_0 < 4.0$ is shown for comparison. Since self-guiding continues to be effective, the phenomenological physics of the LWFA scales very well to higher energies. (bottom) The energy spectrums of self-trapped particles with axis scaled to the appropriate parameters, at laser propagation
 at  distances of 0.5 $L_d$ (left), 0.75 $L_d$ (middle), and 1.0 $L_d$ (right).}
   \label{scaling-self-guiding}
\end{figure}

\begin{figure}[htbp] 
   \centering
   \begin{tabular}{c}
   \includegraphics[width=8.5cm]{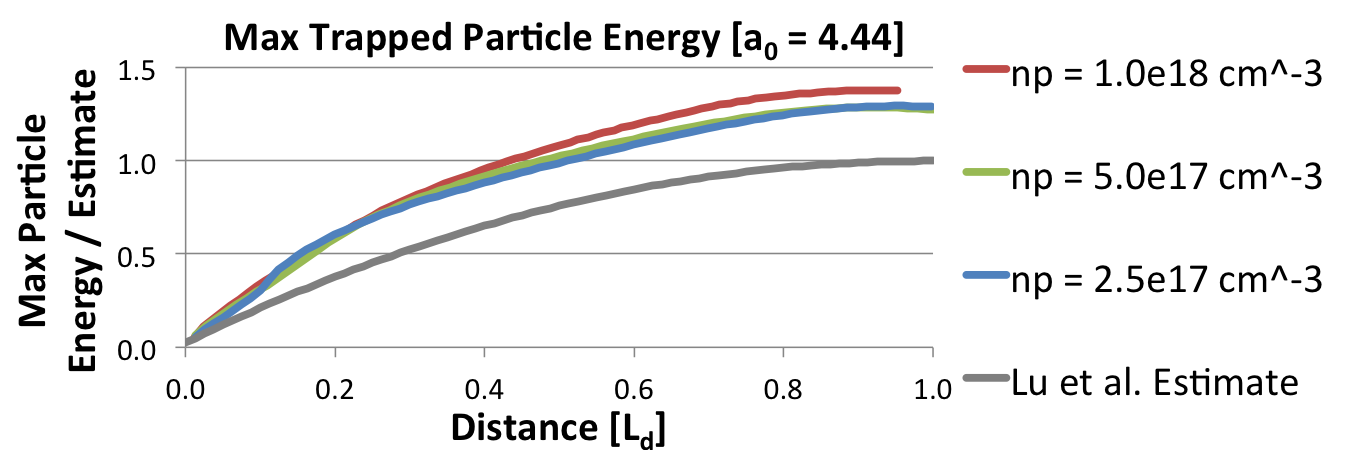} \\
   \includegraphics[width=8.5cm]{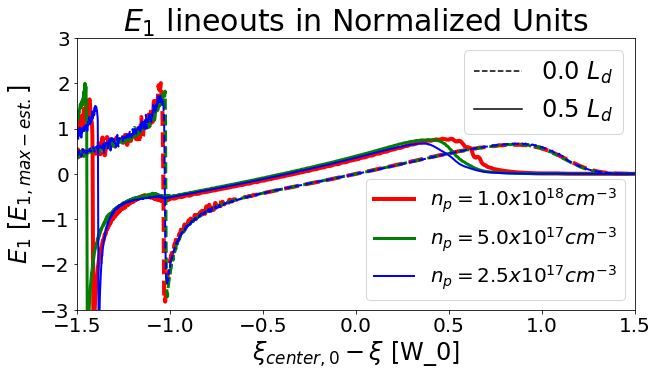}
   \end{tabular}
   \caption{(top) The evolution of the maximum trapped particle energies in normalized units is shown. The result expected from the estimated model in Ref. \cite{weilu} is shown in gray. The reason for the discrepancy in the final value is the downward spike in the accelerating field typical of LWFAs in a cold plasma in the blowout regime, which can be seen in the normalized accelerating field lineouts shown (bottom). The dotted lines represent the lineouts at a laser propagation distance of approximately 0.0 $L_d$, and the solid lines at 0.5 $L_d$. The spike's contribution to the energy is omitted in the Lu et al.'s estimate for simplicity.
   }
   \label{scalea0444-energies}
\end{figure}

\section{Scaling to Higher Trapped Particle Energy}

\begin{figure*}[htbp] 
   \centering
   \begin{tabular}{cc}
   \includegraphics[width=8.5cm]{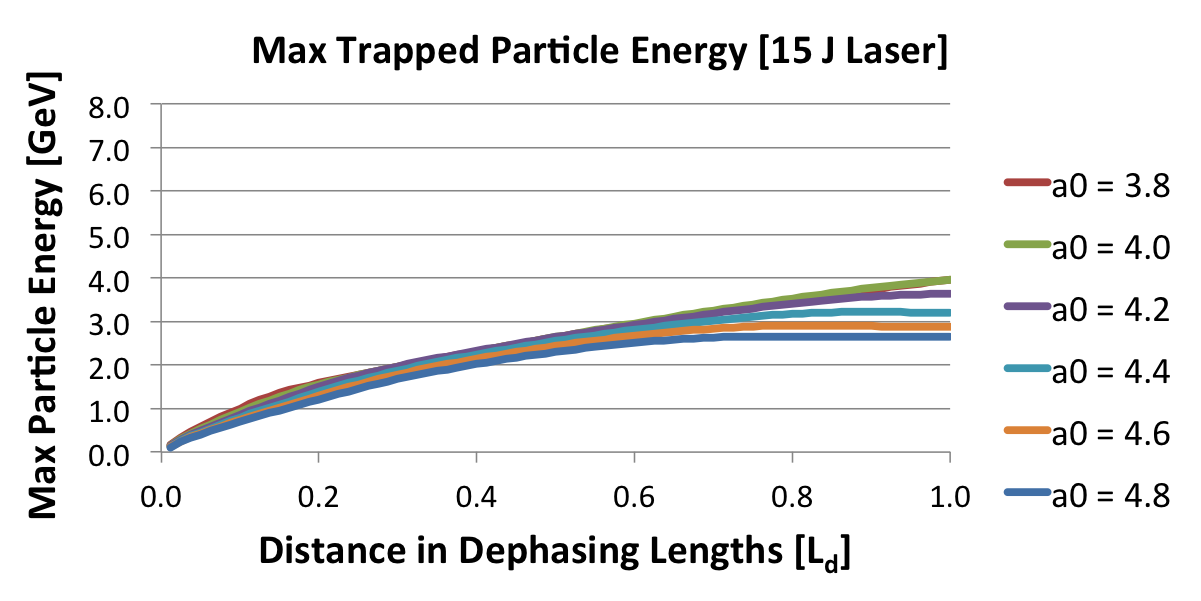}&
     \includegraphics[width=8.5cm]{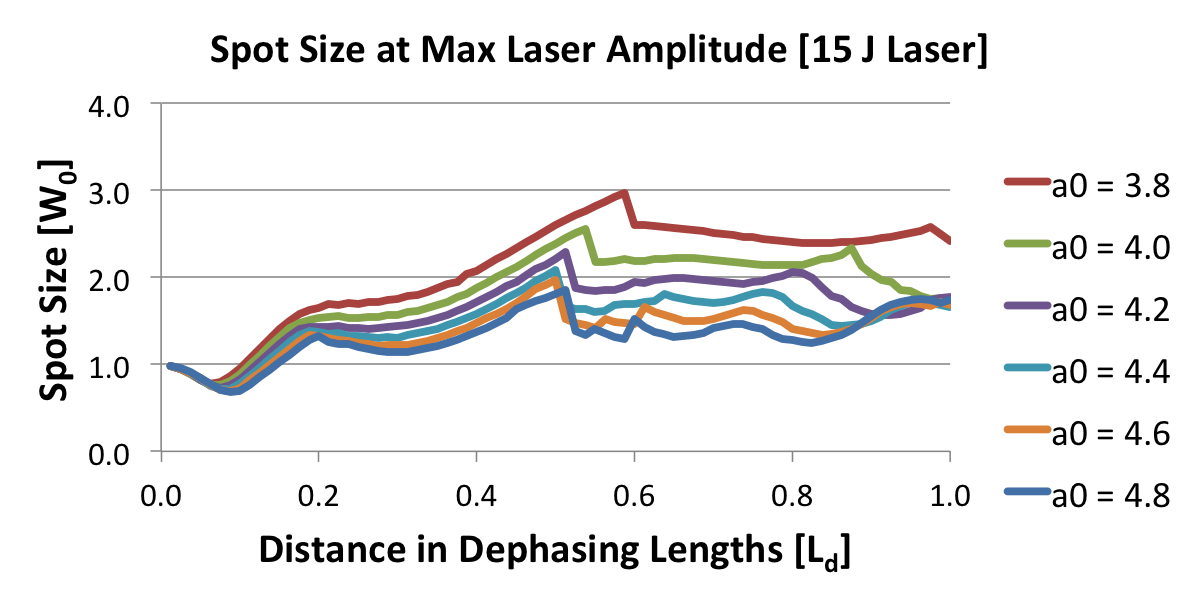} \\
       \includegraphics[width=8.5cm]{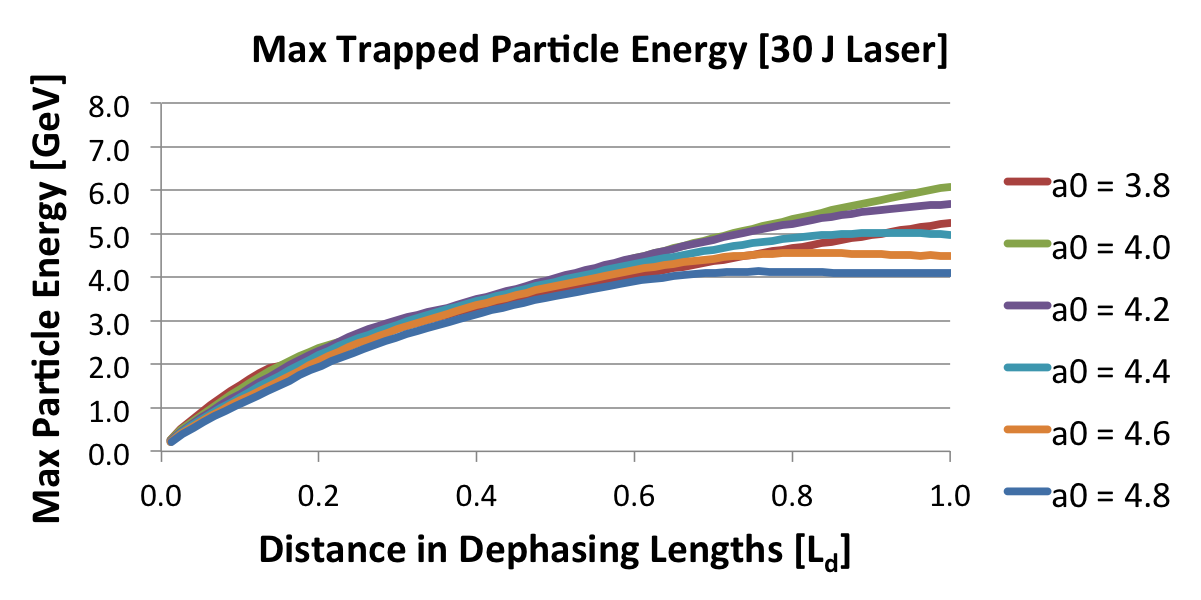} &
  \includegraphics[width=8.5cm]{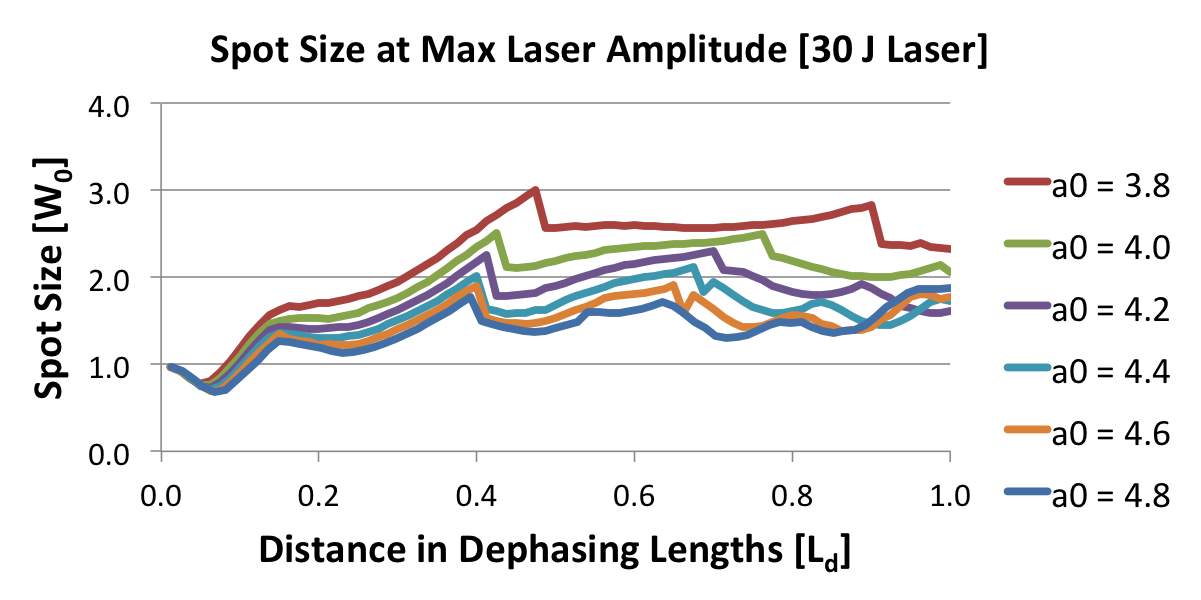}
   \end{tabular}
   \caption{(left) The evolutions of the maximum trapped particle energies is shown for a variety of normalized laser amplitudes ($a_0$), and plotted with the distance traversed by the laser, represented in fractions of the dephasing length, $L_d$. (right) Similarly, the evolutions of the spot sizes, normalized to the initial spot sizes ($W_0$), at the maximum laser amplitude is plotted for varying $a_0$ over these distances. Results are plotted for $15~J$ lasers (top), and $30~J$ lasers (bottom). Here $\tau = \frac{2}{3} W_0$ ($\mathcal{F} = \frac{2}{3}$).}
   \label{a0scale-fixedene}
\end{figure*}

We begin with a set of three LWFA 	quasi-3D OSIRIS simulations with only the m=0 and m=1 modes that illustrate how self-guiding scales to lower densities. The normalized laser amplitude was kept fixed at $a_0 = 4.44$ and the matched spot size and pulse length were scaled from the case used in ref. \cite{weilu} as the density was lowered from $n_p = 1.0 \times 10^{18}$ cm$^{-3}$, to $5.0 \times 10^{17}$ cm$^{-3}$, and finally to $2.5 \times 10^{18}$ cm$^{-3}$. 
The estimated particle energies according to Ref. \cite{weilu} would scale from $2.52$ GeV, $5.28$ GeV, and $10.57$ GeV, respectively. Lu et al. argued that self-guiding would not be as effective as we scale to higher energies if $a_0$ is kept constant, and the acceleration distances in the simulations presented here are $13.8 Z_R$ and $26.4 Z_R$ for the $1.0 \times 10^{18} cm^{-3}$ cm$^{-3}$ and $2.5 \times 10^{-7}$ cm$^{-3}$ densities, respectively. Figure~\ref{densityscaling} demonstrates the qualitative representation of these scaling laws; the structure of the bubble and evolution of trapped particles, in scaled coordinates, appear very similar as we scale to lower plasma density and higher particle energy gain. In the top part of Fig. \ref{scaling-self-guiding} we plot the evolution of the spot size at the location of the maximum laser amplitude in units of initial spot sizes. For these properly matched laser pulses, the evolution of the spot size appears to be very stable, even as the acceleration distance is nearly doubled in Rayleigh lengths. 

These simulations make clear self-guiding does indeed occur to a sufficient extent that the scaling laws given in ref. \cite{weilu} continue to work well as the density is lowered. This is illustrated in the top plot of Fig. \ref{scaling-self-guiding} where the evolution of the laser spots size at the $x_1$ position of the peak intensity is plotted for the three cases as well as for cases with less laser power. Although self-guiding is a phenomenon that does not exactly scale with the rest of the LWFA physics in the blowout regime, it is sufficiently effective throughout the accelerating distance it is evident that other phenomena scale to the appropriate physical units.
In addition, not only do the scaling laws for the wake amplitude, dephasing length, and electron energy hold,  but for fixed laser shapes and amplitudes,  the evolution and shape of the simulation results are also  similar when scaled. This is illustrated by the overlap in the scaled spectrums of the self-trapped particles at the bottom of Fig. \ref{scaling-self-guiding} for the three simulations. Plots are provided at .25, .5, and 1.0 of the normalized accelerating length. The evolution of the maximum trapped particle energy is shown at the top of Fig. \ref{scalea0444-energies} to further illustrate the overlap in the accelerating process when plotted in normalized units. Two of the curves are nearly on top of each other. We also show the prediction of the scaling law from ref. \cite{weilu} gives a lower energy, which is due to a faster energy gain early in time during the simulation that is not included in the estimate. This difference is because for simplicity Lu et al. ignored the role of the  downward spike in the accelerating field (Fig. \ref{scalea0444-energies} bottom) that is common in nonlinear wakes in cold plasmas. This spike (Fig. \ref{scalea0444-energies} bottom), 
provides an extra boost in the early stages of particle acceleration, resulting in a higher final energy (note that the slope of the curves become similar after about .4 $L_d$. We note that the phenomenological model and our observations of scaling applies to externally injected beams as well, and the beams here are self-injected only as an example. 

\begin{figure*}[htbp] 
   \centering
   \begin{tabular}{cc}
   \includegraphics[width=8.5cm]{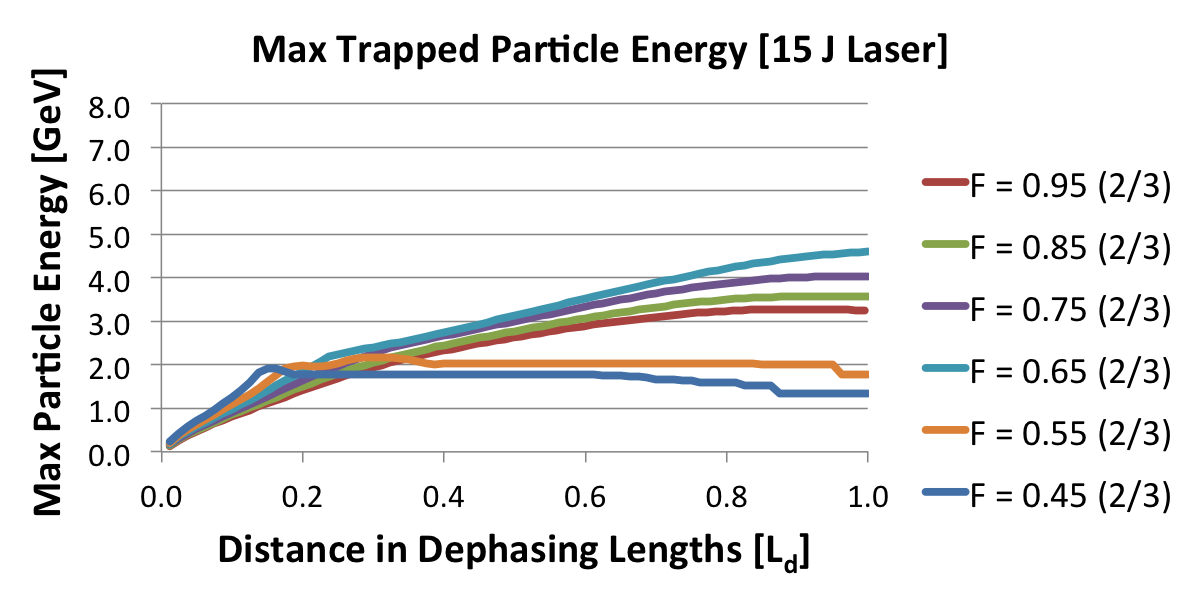} &
    \includegraphics[width=8.5cm]{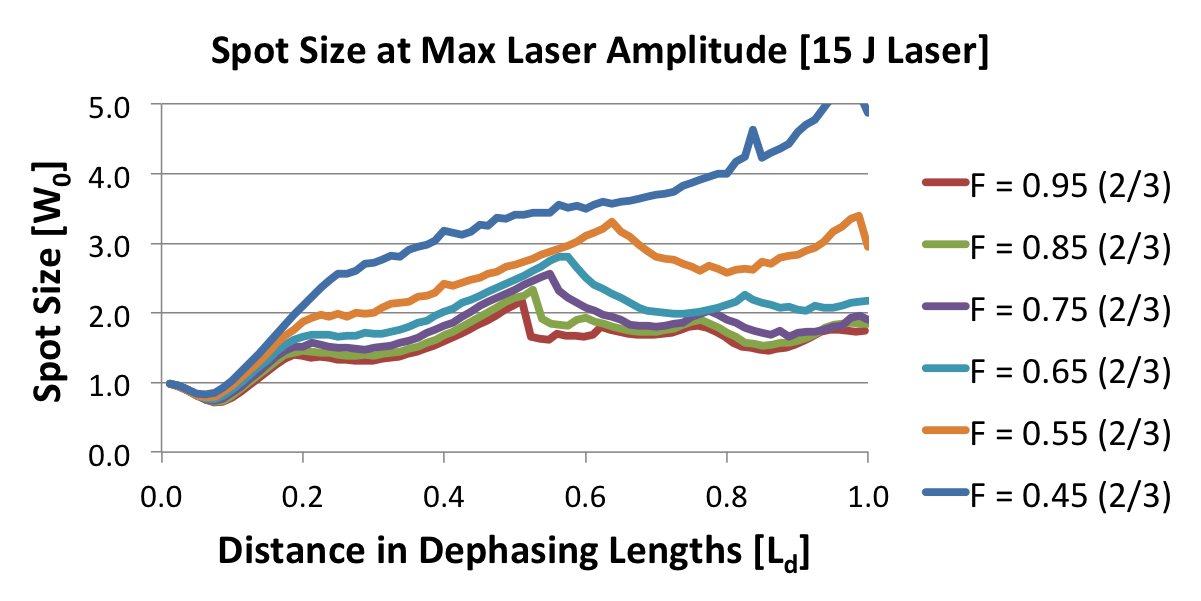} \\
      \includegraphics[width=8.5cm]{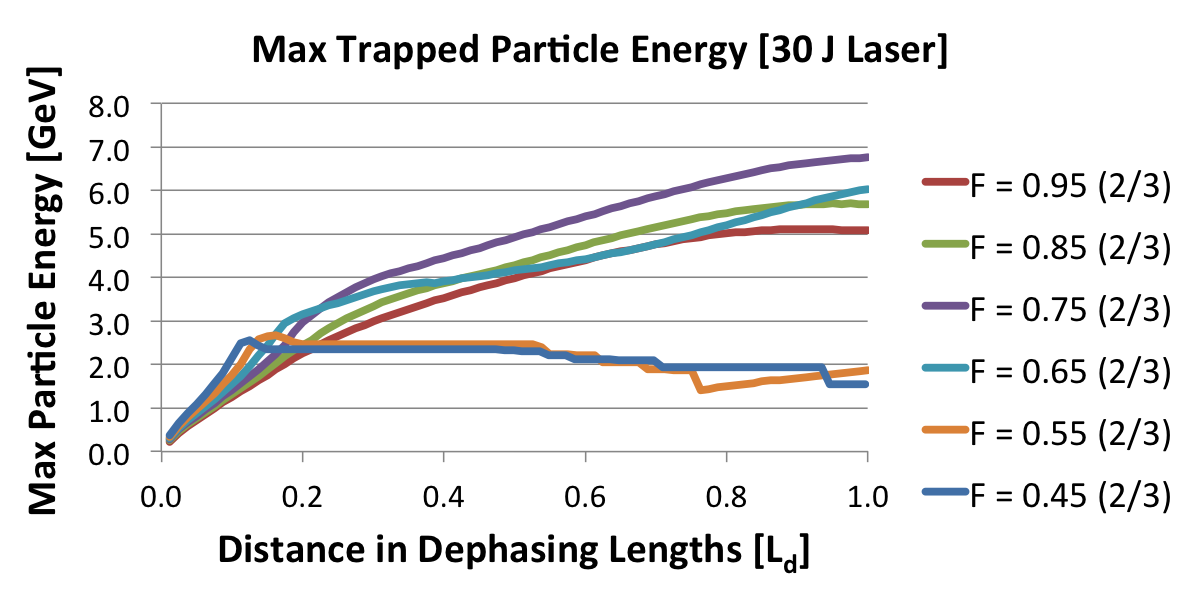} &
    \includegraphics[width=8.5cm]{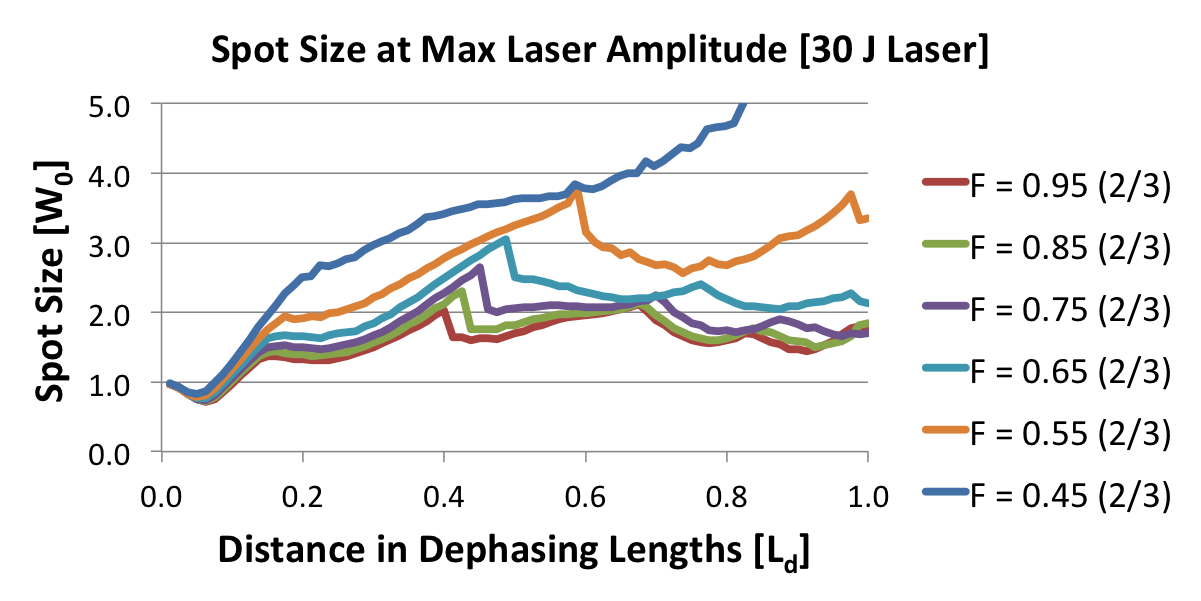} \\
    \includegraphics[width=8.5cm]{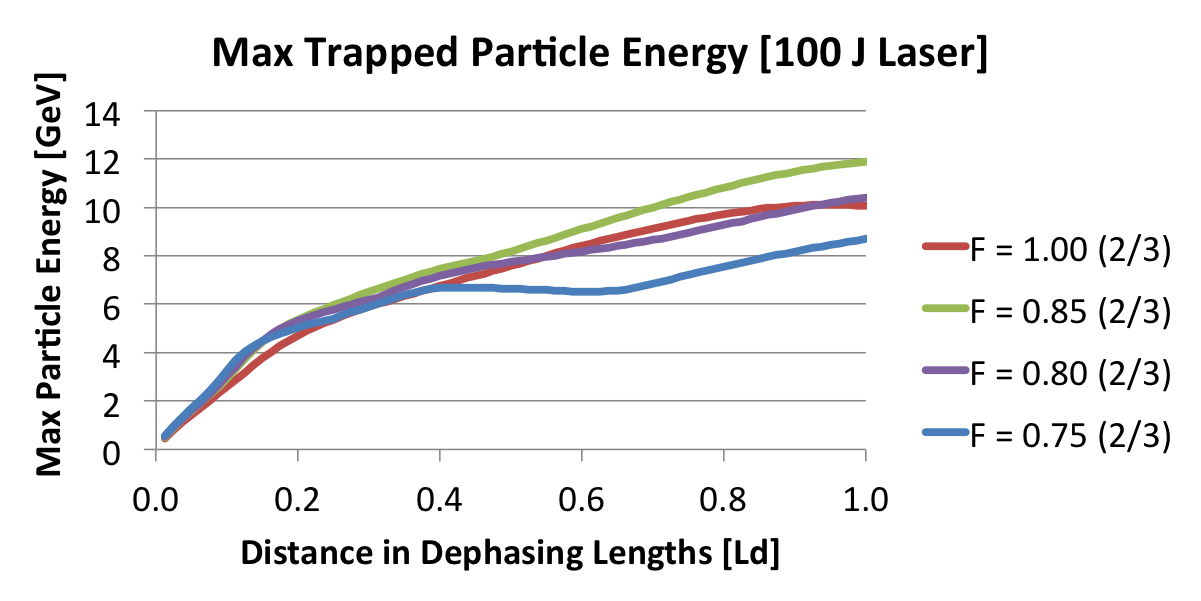} &
    \includegraphics[width=8.5cm]{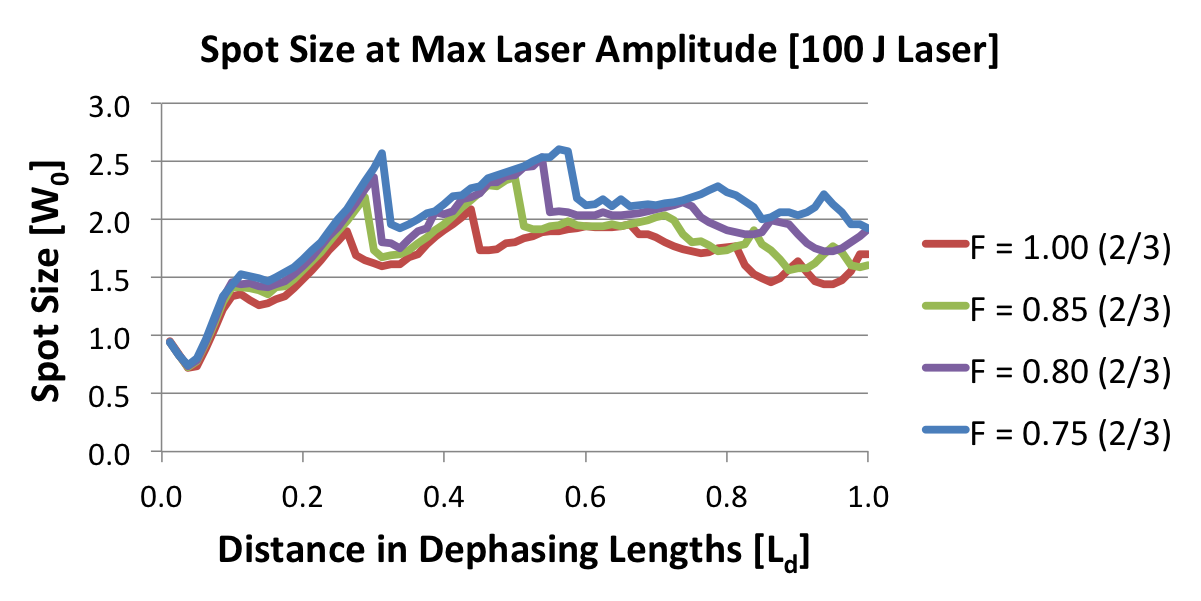} \\
   \end{tabular}
   \caption{(left) The evolutions of the maximum trapped particle energies is shown for a variety of normalized pulse lengths ($\mathcal{F}$), and plotted with the distance traversed by the laser, represented in fractions of the dephasing length, $L_d$. (right) Similarly, the evolutions of the spot sizes, normalized to the initial spot sizes ($W_0$), at the maximum laser amplitude is plotted for varying $\mathcal{F}$ over these distances. Results are shown for lasers with a total energy of $15~J$ (top), $30~J$ (middle), and $100~J$ (bottom).}
   \label{fwhm-energies-1530J}
\end{figure*}

\section{Scaling with Fixed Laser Energy}

In all simulations presented this paper, the spot sizes were matched for self-guiding. Unless otherwise stated, the normalized laser amplitude is $a_0 = 4.44$. These simulations were conducted over their respective estimated dephasing lengths, $L_d$, fwith a cell resolution of $\Delta z k_0 = 0.2$, and $\Delta r k_p \approx 0.1$ for the $15$ J and $30$ J simulations, and $\Delta r k_p \approx 0.2$ for the $100$ J simulations. The ($r$,$z$) box sizes ranged from $(6.4 W_0,7.6W_0)$ for the $15$ J, $\mathcal{F} = 0.63$ simulation and $(4.5W_0,17W_0)$ for the $100$ J, $\mathcal{F} = 0.5$ simulation, and were appropriately increased to ensure the boundary effect did not affect the results over many Rayleigh lengths.  $2$ particles were initialized in the $z$ direction of each cell, and $8$ particles were initialized along $\phi$. The simulations were performed in the quasi-3D geometry with the azimuthal modal decomposition truncated at $m = 1$.

\subsection{Scaling Theory}

First we discuss the theory with which we examine the various optimization methods we will explore in the following sections. With respect to experiments and with optimizing a laser pulse shape, it is important to investigate what can be done with a fixed laser energy. In previous sections have established that self-guiding is effective even with a fixed $a_0$ as we scale to a beam energy past $10$ GeV. Considering that sufficient self-guiding is a foundational assumption that these scalings depend on, and that this assumption is satisfied up to this energy, we have a basis for further examining these scalings for a fixed energy. We keep in mind that the scaling laws give a scaling of the energy gain as $a_0 \frac{\omega_0^2}{\omega_p^2}$. 
The total laser energy $E_L$ can be calculated by the laser power and pulse length as
\begin{equation}
\label{elestimate}
E_L = \alpha P \tau,
\end{equation}
where $\tau \equiv \tau_{\text{FWHM}}$, and $\alpha$ is a constant that depends on the exact shape of the longitudinal profile  (for the longitudinal polynomial profile we used for the simulations in this paper, $\alpha \approx 1.04365$). 
We will assume that the pulse length is some specified fraction $\mathcal{F}$ of the spot size $W_0$, or $\tau = \mathcal{F} W_0$. Lu et al.\cite{weilu} matches the estimated pump depletion distance $L_p$ and the dephasing length $L_d$ by setting $\mathcal{F} = 2/3$ when the spot size is matched, but we may find empirically that there is a better choice for this value. In combination with the matched spot size condition. We now rewrite the energy of the laser in terms combinations of parameters that are useful when calculating the energy gain
\begin{equation}
\label{taunp}
E_L= \frac{\alpha}{4} \left(a_0 \frac{\omega_0^2}{\omega_p^2}\right)^{2/3}\frac{a_0^2\mathcal{F}}{\omega_0}
\end{equation}
With respect to optimizing the pulse length we will let, $\mathcal{F}$, be a free parameter. 
We can now use this expression to rewrite the energy gain in terms of $E_L$, $a_0$, $\mathcal{F}$, and $\omega_0$,   where it was expressed in Ref. \cite{weilu} as a function of laser power and plasma density. The result is 
\begin{equation}
\label{matchedestene}
\Delta E = \frac{2}{3} \frac{m_e c^2}{\alpha^{2/3}} \left[ \frac{4 \omega_0}{\mathcal{A}}\right]^{2/3} \frac{E_L^{2/3}}{\mathcal{F}^{2/3} a_0^{4/3}},
\end{equation}
where $\mathcal{A}$ is a constant that is equal to $17~GW$ in MKS units. This equation expresses the fact that, if you reduce the pulse length $\mathcal{F}$ for a fixed $E_L$ and $a_0$, you are effectively widening the pulse width $W_0$ to compensate. In order to keep the the spot size matched this requires that the density be lowered. This causes the laser to be matched to a lower density plasma, which would give the LWFA a longer acceleration length and a higher overall particle energy. Interestingly,  the matched spot size could also be increased by increasing $a_0$, however this would lead to 
a lower beam energy overall. 

\subsection{Choice of  $a_0$}

First we examine the effect of changing $a_0$ and discuss our choice of a particular default value of $a_0 = 4.44$. Optimizing the particle energy by reducing $a_0$ is limited by the fact that the laser power needs to exceed ta threshold for effective self-guiding. This has been found empirically to be $a_0 \gtrsim 4.0$ is required\cite{weilu,weilu2} as was demonstrated in Fig. \ref{scaling-self-guiding} (together with the fact that self-guiding fails at low $a_0$). We have examined self-guiding in more detail for the subtler effects of adjusting $a_0$ to a variety of values near $4.0$ while keeping the laser energy fixed at $15~J$, $30~J$, and $100~J$. The parameters explored during this process is shown in Table~\ref{Scana01530J} and the results of the simulations are shown in Fig. \ref{a0scale-fixedene}. We find from the energy evolutions plotted on the left side of Figure~\ref{a0scale-fixedene} that indeed the maximum energy gain improves in general with lowered $a_0$, but from the spot size evolution shown on the right side we can see that the stability of the spot size (although still self-guided) is affected below $a_0 = 4.4$. On the bottom left plot for a $30~J$ laser we see that for $a_0 = 3.8$ the parameters have just reached the threshold of failure for self-guiding, and the final energy is lower than the $a_0 = 4.0$ case, despite the predictions of the scaling theory. The scaling theory assumes that the accelerating field structure is stable throughout the dephasing length, and when self-guiding fails before the end of the total distance we achieve a lower overall accelerated energy. 

As we will discuss in the next section, the stability of the wake is also affected by the reduction of the normalized pulse length, $\mathcal{F}$. Adjusting $\mathcal{F}$, however, leads to a greater advantage in the gain in the accelerated energy due to the scalings represented in Equation~\ref{matchedestene}. We found that setting $a_0 = 4.44$ is effective to ensure a stable spot size evolution without unnecessary reduction of the particle energy. 

\begin{figure}[t] 
   \centering
   \begin{tabular}{l}
      \includegraphics[width=8.0cm]{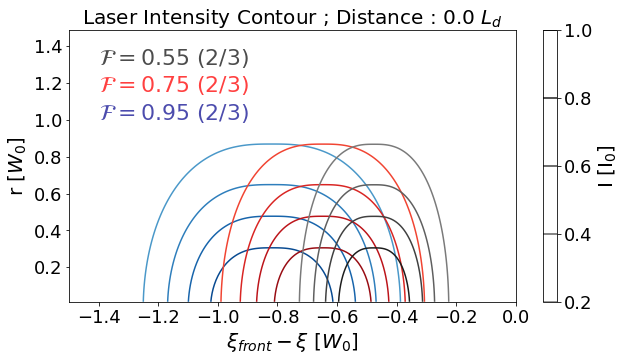} \\
      \includegraphics[width=8.0cm]{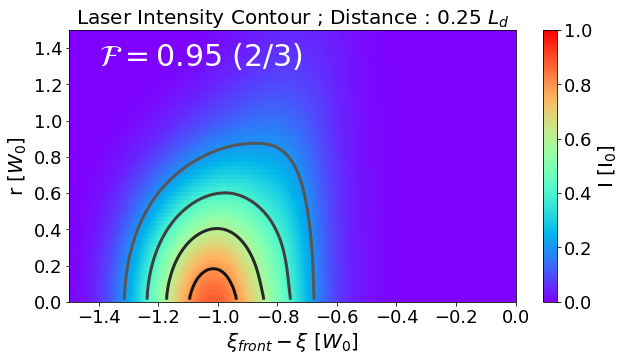} \\
      \includegraphics[width=8.0cm]{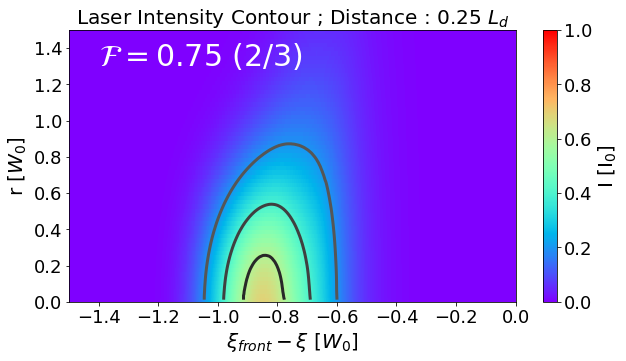} \\ 
      \includegraphics[width=8.0cm]{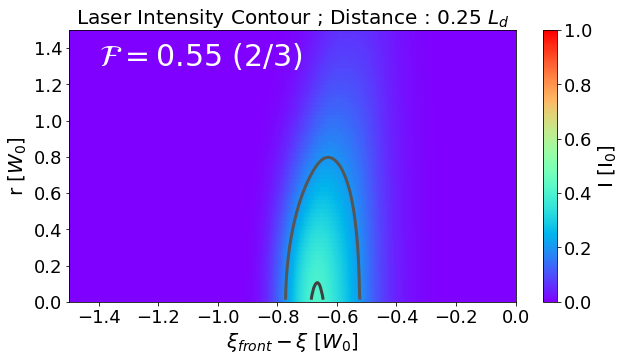}
   \end{tabular}
   \caption{Contrasted contour plots of the initial, normalized, laser profile is shown (top), as well as the evolved contours in each case at 0.25 $L_d$ (bottom three).}
   \label{contours}
\end{figure}

\subsection{Choice of Pulse Length}

\begin{table*}[htbp]
   \centering
   \begin{tabularx}{\textwidth}{c@{\extracolsep{\fill}}cccccccccc } 
      \toprule
      \multicolumn{8}{c}{Calculated } & \vline & \multicolumn{2}{c}{Simulated  } \\
      \hline 
      a0 & P &  $\tau$ & $n_p$ & $Z_R$ & $W_0$ & $L_d$  & Est. E &\vline & Q & Max E \\
       & (TW) & (fs) & ($cm^{-3}$) & ($cm$) & ($\mu m$) & $(cm)$ & ($GeV$) & \vline &($pC$) & ($GeV$) \\
      \hline
      \multicolumn{11}{c}{$15J$ Laser} \\
      \hline
       3.80 & 271  & 53.1  & 7.52 & 0.224 & 23.9 & 3.70 & 3.01 & \vline & 231 & 3.97  \\
       4.00 & 280  & 51.3  & 8.48 & 0.210 & 23.1 & 3.17 & 2.81 & \vline & 176 &  3.97 \\
       4.20 & 289  & 49.7  & 9.50 & 0.196 & 22.4 & 2.74 & 2.63 & \vline & 265 &  3.64 \\
       4.40 & 298  & 48.1  & 10.6 & 0.185 & 21.7 & 2.38 & 2.47 & \vline & 337 &  3.22 \\
       4.60 & 307  & 46.8  & 11.7 & 0.174 & 21.0 & 2.08 & 2.33 & \vline & 402 &  2.91 \\
       4.80 & 316  & 45.5  & 13.0 & 0.164 & 20.5 & 2.08 & 2.20 & \vline & 477 &  2.66 \\
      \hline
      \multicolumn{11}{c}{$30J$ Laser} \\
      \hline
       3.80 & 429  & 66.9  & 4.74 & 0.356 & 30.1 & 7.40 & 4.77 & \vline & 64.1 &  5.26 \\
       4.00 & 444  & 64.7  & 5.34 & 0.333 & 29.1 & 6.34 & 4.46 & \vline & 148  &  6.08 \\
       4.20 & 459  & 62.6  & 5.98 & 0.312 & 28.2 & 5.48 & 4.18 & \vline & 225  &  5.69 \\
       4.40 & 473  & 60.7  & 6.67 & 0.293 & 27.3 & 4.76 & 3.93 & \vline & 317  &  5.02 \\
       4.60 & 487  & 58.9  & 7.40 & 0.276 & 26.5 & 4.17 & 3.70 & \vline & 401  &  4.56 \\
       4.80 & 502  & 57.3  & 8.17 & 0.261 & 25.8 & 3.67 & 3.50 & \vline & 506  &  4.13 \\
   \end{tabularx}
   \caption{These are the parameters of LWFA simulations for $15~J$ and $30~J$ lasers, given an optimal density. The ratio of pulse length to spot size is such that $\tau = \frac{2}{3} W_0$. The charge and the max energy of the mono-energetic particle beam are also shown.}
   \label{Scana01530J}
\end{table*}
\begin{table*}[htbp]
   \centering
      \begin{tabularx}{\textwidth}{c@{\extracolsep{\fill}}cccccccccc} 
      \toprule
      \multicolumn{8}{c}{Calculated } & \vline & \multicolumn{2}{c}{Simulated  } \\
      \hline
      $\mathcal{F}$ & P &  $\tau$ & $n_p$ & $Z_R$ & $W_0$ & $L_d$  & Est. E & \vline & Q & Max E \\
      \% $\frac{2}{3}$ & (TW) & (fs) & ($cm^{-3}$) & ($cm$) & ($\mu m$) & $(cm)$ & ($GeV$) &\vline &($pC$) & ($GeV$) \\
      \hline
      \multicolumn{11}{c}{$15J$ Laser} \\
      \hline
       95 & 324  & 46.0  & 10.5 & 0.188 & 21.9 & 2.43 & 2.52 &\vline & 355  & 3.25  \\
       85 & 348  & 42.8  & 9.75 & 0.202 & 22.7 & 2.71 & 2.71 &\vline & 301  & 3.57  \\
       75 & 378  & 39.4  & 8.95 & 0.220 & 23.7 & 3.08 & 2.95 &\vline & 227  & 4.04  \\
       65 & 416  & 35.8  & 8.15 & 0.242 & 24.8 & 3.54 & 3.24 &\vline & 83.4 & 4.60  \\
       55 & 464  & 32.1  & 7.28 & 0.271 & 26.3 & 4.20 & 3.63 &\vline & N/A & N/A  \\
       45 & 532  & 28.0  & 6.38 & 0.309 & 28.1 & 5.11 & 4.14 &\vline & N/A & N/A  \\
      \hline
      \multicolumn{11}{c}{$30J$ Laser} \\
      \hline
       95 & 513  & 58.2  & 6.60 & 0.299 & 27.6 & 4.86 & 4.00 &\vline & 320  & 5.08  \\
       85 & 554  & 54.0  & 6.13 & 0.322 & 28.6 & 5.43 & 4.31 &\vline & 235  & 5.69  \\
       75 & 602  & 49.7  & 5.64 & 0.350 & 29.8 & 6.15 & 4.68 &\vline & 103  & 6.76  \\
       65 & 662  & 45.2  & 5.12 & 0.385 & 31.3 & 7.11 & 5.16 &\vline & 2.99  & 6.04    \\
       55 & 738  & 40.5  & 4.58 & 0.431 & 33.1 & 8.41 & 5.77 &\vline & N/A  & N/A    \\
       45 & 845  & 35.4  & 4.01 & 0.492 & 35.4 & 10.3 & 6.59 &\vline & N/A  & N/A     \\
      \hline
      \multicolumn{11}{c}{$100J$ Laser} \\
      \hline
       100 & 1063  & 90.1  & 3.05e17& 0.646 & 40.6 & 15.5 & 8.66 &\vline  & 227 &10.1  \\
       85 & 1185  & 80.9  & 2.74e17 & 0.721 & 42.8 & 18.2 & 9.65 &\vline & 107 & 11.9  \\
       80 & 1233  & 77.7  & 2.63e17 & 0.750 & 43.7 & 19.3 & 10.0 &\vline & N/A & 10.4  \\
       75 & 1288  & 74.4  & 2.52e17 & 0.783 & 44.6 & 20.6 & 10.5 &\vline & N/A &  8.70  \\
   \end{tabularx}
   \caption{These are the parameters of LWFA simulations for $15~J$, $30~J$, and $100~J$ lasers, given an optimal plasma density in each case for the greatest energy gain. The initial amplitude of the laser is kept at $a_0 = 4.44$. The charge of the mono-energetic particle beams are shown (if no mono-energetic feature, the charge is N/A). }
   \label{Scanf1530100J}
\end{table*}

A variety of simulations were conducted to explore various normalized pulse lengths, $\mathcal{F}$, for a 15 J, 30 J, and 100 J laser, given the chosen amplitude of $a_0 = 4.44$. The parameters for these simulations are displayed in Table~\ref{Scanf1530100J}. We have found that a reduction of $\mathcal{F}$ is an effective way of optimizing the maximum energy gain of the trapped particles, granted that the self-guiding mechanism of the LWFA is not significantly affected.

\begin{figure}[htbp] 
   \centering
   \begin{tabular}{c}
   \includegraphics[width=8.5cm]{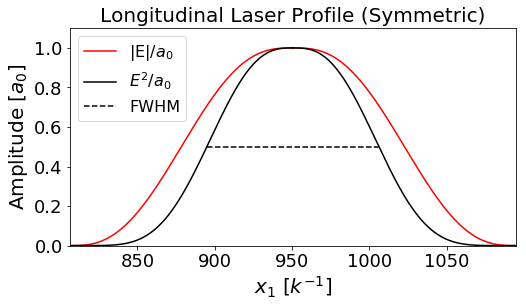} \\
   \includegraphics[width=8.5cm]{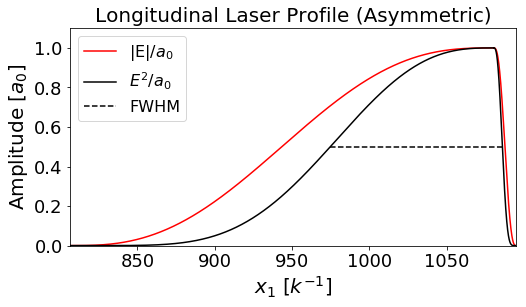}  
   \end{tabular}
   \caption{Illustration for symmetric and asymmetric axial laser profiles. (bottom) An example of a longitudinally asymmetric laser profile that we used for the 15 J, $\mathcal{F} = 0.95 (2/3)$ simulation, with a 5\% rise time as compared to the total rise and fall times. (top) A symmetric profile with a 50\% rise time for comparison.}
   \label{asym-laser-profile}
\end{figure}

\begin{figure*}[htbp] 
   \centering
   \begin{tabular}{cc}
   \includegraphics[width=8.5cm]{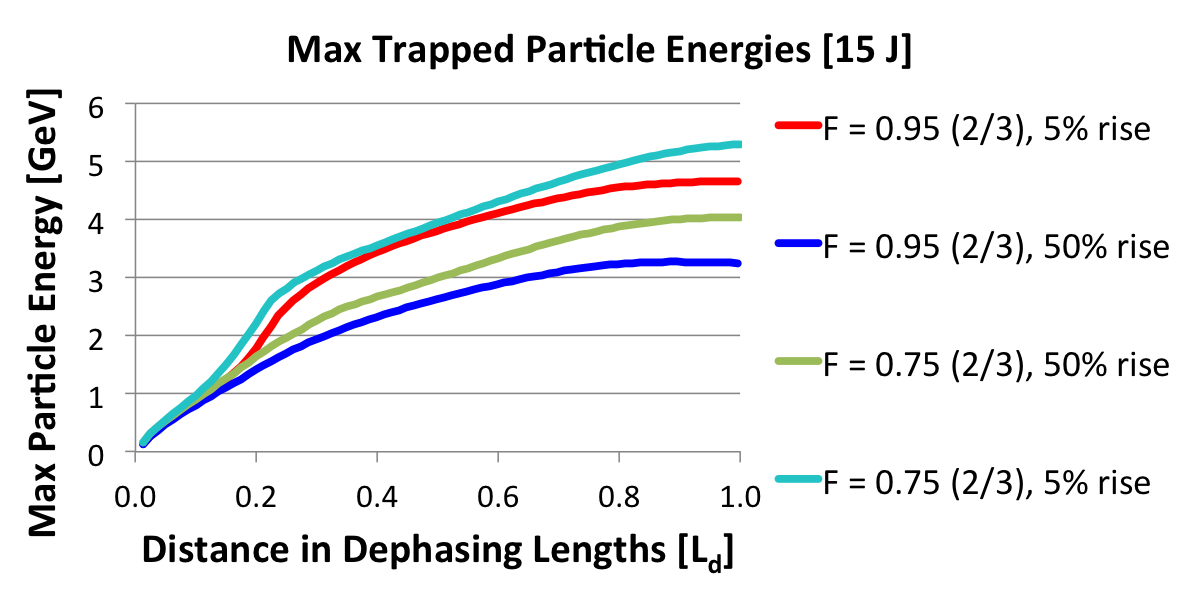} &  \includegraphics[width=8.5cm]{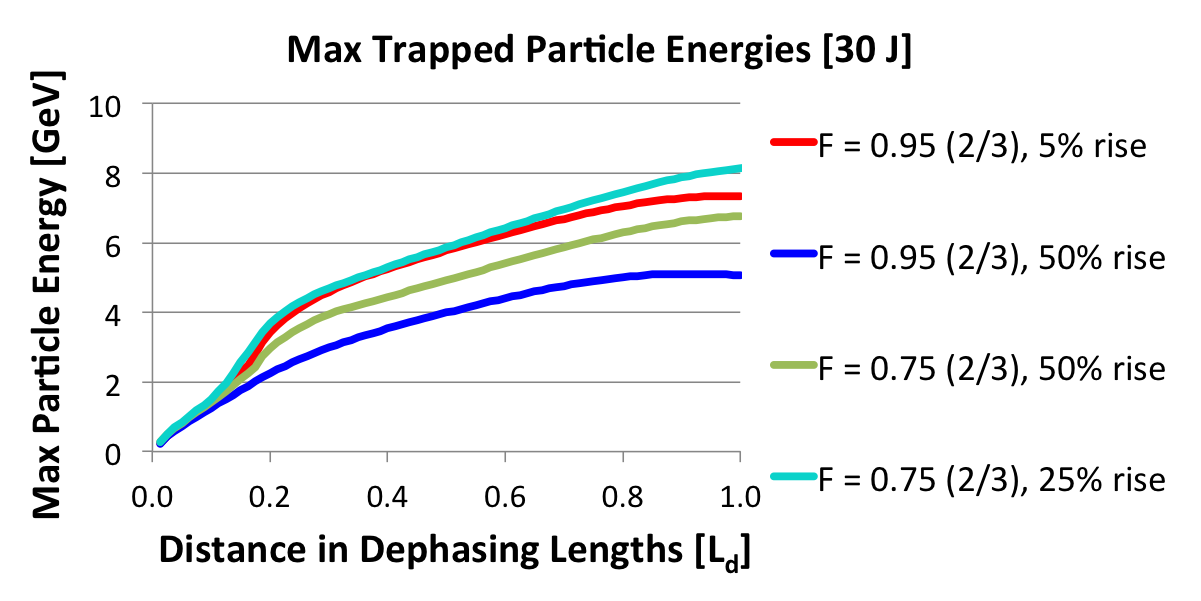}\\
\includegraphics[width=8.5cm]{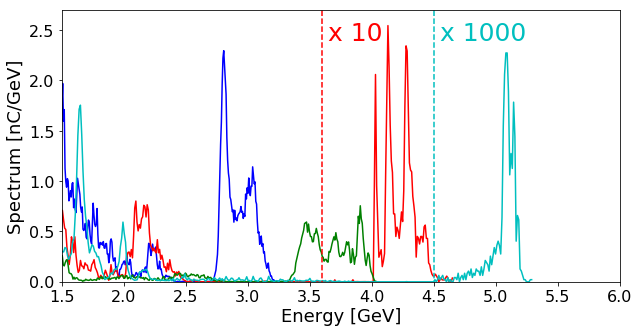} & \includegraphics[width=8.5cm]{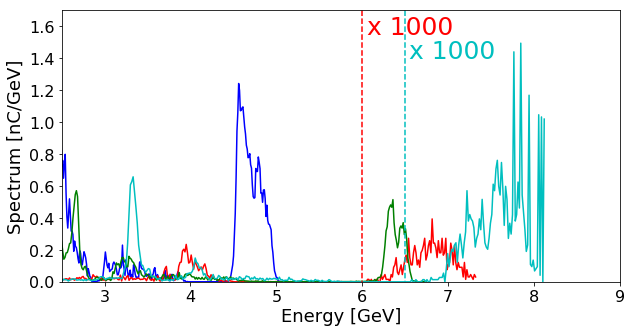}
   \end{tabular}
   \caption{(top) The evolution of the particle beam energies for various normalized pulse lengths and proportional rise times. (bottom) The energy spectrums of the trapped particles, after laser traversing 1.0 $L_d$. Results are shown for $15~J$ lasers (left) and $30~J$ lasers (right). Note that on the bottom left plot the y-axis value for the red line is multiplied by $10$ to the right of the vertical, dotted red line, and the cyan line is multiplied by $1000$ to the right of the dotted cyan line. On the bottom right plot the red and cyan line is multiplied by $1000$ on the right side of their respective vertical dotted lines. 
   }
   \label{fwhm-energies-1530J-risescan}
\end{figure*}

In Ref.~\cite{Decker96} it was explained that self-guiding is possible despite the leading edge of the laser diffracting because the laser is continuously being pump-depleted before it can diffract. The undepleted potion of the laser, which exists behind the density compression formed locally by the leading edge the laser, remains guided. There is a lower limit to which one may reduce the normalized pulse length before, proportionally, the `leading-edge' of the laser forms the bulk of the laser's length, causing self-guiding to fail. This is evidenced in the left ride of Fig. \ref{fwhm-energies-1530J} where the maximum accelerated particle energy at the end of each scaled run abruptly drops at some threshold $\mathcal{F}$. The optimal normalized pulse length is higher for higher laser energies because the number of Rayleigh lengths over the acceleration distance increases, resulting in a greater portion of the laser front being diffracted. There is an optimal pulse length at which the pump depletion process competes sufficiently with the diffraction process in order for self-guiding to be stable, and the LWFA is able to accelerate particles to a higher energy for that given laser energy. Beyond this point the self-guiding is no longer stable, as indicated by the bottom of Fig.~\ref{contours} where the contours of the laser is shown at .25 $L_d$ for different pulse lengths. In Fig.~\ref{contours}  is can be seen that for $\mathcal{F}$ =.55 (2/3) the laser is too short to be guided, as the bulk of the laser profile has already broadened due to diffraction. In the center images of \ref{contours}, however, it can be seen that the distinctly guided back-half of the laser retains its spot size, while the somewhat pump-depleted front-end is beginning to diffract. In Fig.~\ref{fwhm-energies-1530J} and in Table~\ref{Scanf1530100J} we show how the energy and charge of a self-injected quasi-monoenergetic bunch changes with pulse length. By reducing the normalized pulse length to $65 \%$ of the default length (of $\frac{2}{3}W_0$) in the 15 J case, we were able to increase the maximum particle energy from 3.25 GeV to 4.00 GeV,
with a loss in the total charge of the self-trapped, quasi-mono-energetic bunch from 355 $pC$ to 83.4 $pC$. The 30 J and 100 J cases showed best results at $75\%$ and $85 \%$ the default pulse lengths, respectively, with an energy increase from 5.08 GeV (320 $pC$) to 6.76 GeV (103 $pC$), and from 10.1 GeV (227 $pC$) to 11.9 GeV (107 $pC$), respectively. It is important to note that the accelerating mechanism is separate from the trapping mechanism, and this loss in charge may be avoided or compensated for through a different injection scheme. However, it is clearly demonstrated that the maximum particle energy attainable for a laser of a given energy may be improved in this way.

\section{Longitudinal Pulse Shape}

Next, we explore further optimization by adjusting the longitudinal profile for a fixed laser energy. Tzoufras et al.\cite{tzoufras14} showed that by tailoring the longitudinal profile and applying a low-amplitude pre-pulse, it is possible to improve the stability and efficiency of an LWFA. They matched the longitudinal profile to the equilibrium profile which it evolves to late in the simulation and considered higher laser intensities. With our newly implemented quasi-3D geometry, it is now possible to explore a variety of longitudinal profiles over a range of parameters quickly. 

Here, we explore profiles which do not have a pre-pulse but are forwardly skewed (with fixed laser energy) and examine how the electron energy is effected. Simulations presented so far have implemented a symmetric profile, where the pulse rise (position of max amplitude to the front) is equal to the pulse fall (max amplitude to the back), which we define as a `$50\%$ rise'. Fig. \ref{fwhm-energies-1530J-risescan} presents the beam energy gain for simulations with a $5\%$ rise but with the fwhm and peak intensity being kept fixed to previous values (see Fig.~\ref{asym-laser-profile}). 

We found that often a forwardly skewed pulse generated particles of higher energy. This is because as the front of the pulse depletes, the maximum laser amplitude and the bubble radius slowly shrink. The downward spike in the accelerating field thus evolves in phase with the trapped particles at the beginning of the simulation, increasing the overall beam energy in a reverse accordion-like effect\cite{Katsouleas86}. As can be seen if Fig.~\ref{fwhm-energies-1530J-risescan}, for the 15 $J$ case, a max particle energy of 4.66 $GeV$ (39.4 $pC$ bunch) was achieved with a $5\%$ rise as compared to 3.25 $GeV$ (355 $pC$ bunch) for $\mathcal{F} = 0.95 (2/3)$. For the $30~J$ laser case we found that particle energies of 7.34 $GeV$ (0.14 $pC$) was achieved with the same $5\%$ rise time as compared to 5.08 $GeV$ (320 $pC$). As per the pulse length optimization, this longitudinal profile optimization results in a loss of the total charge of self-trapped particles, but the total charge may be mediated either by externally injecting particles rather than relying on self-injection as we did in this case. 

In addition, we found that for the $15~J$ laser a higher total energy gain was found when we simultaneously shortened the normalized pulse length as we implemented a 5\% rise time, where the every was 5.3 $GeV$ (0.26 $pC$) as compared to $4 GeV$ (227.0 $pC$) for a symmetric pulse with $\mathcal{F} = 0.75 (2/3)$ (see Table~\ref{AsymScan1530J}). As can be seen, this method of increasing the final particle energy through asymmetric profiles may be combined with the method of increase through longitudinal pulse length shortage discussed in the previous section. Table~\ref{AsymScan1530J} and Figure~\ref{fwhm-energies-1530J-risescan} show examples of cases in which we combined an asymmetric pulse shape with a reduced pulse length. The best case we have examined so far for the $30~J$ laser involve a $25\%$ rise time and $\mathcal{F} = 0.75 (2/3)$, producing energies as high as 8.15 $GeV$ (0.47 $pC$) as compared to 6.76 $GeV$ (103 $pC$) of the equivalent simulation with a symmetric pulse shape. For the $30~J$, pulse length optimized ($\mathcal{F}$ = 0.75) case we started with simulating a $5 \%$ rise as in the $15~J$ case and found the change in the energy gain to be small (up to 7.07 $GeV$ as compared to 6.76 $GeV$), and therefore examined longer asymmetric rise times to obtain a better energy gain. For higher laser energies the ratio of the Rayleigh to the dephasing length is greater, and therefore a proportionally greater amount of the front of the laser diffracts at a different rate. In order to improve this optimization method one must find a laser profile with which the accelerating bubble shape and structure evolve to apply the greater field on the trapped particles for a greater, extended amount of time. Therefore, the optimal longitudinal profile is different for laser pulses of higher energies.

The results show that we do in fact get a combined gain in the final maximum particle energy, but with a significant loss in the final trapped bunch charge (bringing it to the order of a fraction of $pC$). However, the accelerating mechanism is separate from the particle trapping mechanism, and this loss in charge may be avoided or compensated for through an alternative particle trapping scheme. Only the simplest case of this optimization method is presented here to discuss its mechanism and potentiality. The accelerating mechanism appears to show it is possible to increase the maximum particle energy of a LWFA with a 15 $J$ laser to 5.3 $GeV$, which is more than double the traditional, unoptimized Lu et al. estimate of 2.52 $GeV$. 

Overall, this simulations show that, given a laser with a fixed energy, it is possible to optimize these parameters to acquire trapped particle characteristics ideal for a desired future application. Future work may involve using more controlled self-injection methods combined with better tuning the re-phasing mechanism of the spike to reduce the loss in the total accelerated charge as well as considering more complicated laser profiles, while further exploring the full potential of combining various energy gain optimization methods. 

\begin{table}[htbp]
   \centering
   \begin{tabular}{@{} ccccccccc @{}} 
      \toprule
      $\mathcal{F}$ & Rise Time  & Q & Max E & \vline & $\mathcal{F}$ & Rise Time  & Q & Max E\\
       \% $\frac{2}{3}$ & \%  & ($pC$)  & ($GeV$) & \vline & \% $\frac{2}{3}$ & \%  & ($pC$)  & ($GeV$) \\
      \hline
       \multicolumn{4}{c}{$15J$ Laser } & \vline & \multicolumn{4}{c}{$30J$ Laser } \\
      \hline
        95 &  50  & 355 & 3.25 & \vline & 95 &  50 & 320 &  5.08 \\
        75 &  50  & 227  & 4.04 & \vline  & 75 &  50 &  103 & 6.76\\
        95 &  5  & 39.4 & 4.66  & \vline & 95 &  5 & 0.14 & 7.34 \\
        75 &  5  & 0.26 & 5.30 & \vline & 75 & 5 & N/A & 7.07\\
         & & & & \vline & 75 & 15 & N/A & 7.52\\
        & & & & \vline &  75 &  25 &  0.47 &  8.15 \\
   \end{tabular}
   \caption{These are the parameters for a series of LWFA simulations with pulse-length optimization, asymmetric longitudinal profile lasers, and combinations of the two. The charges presented are the total charges of the quasi-mono-energetic high-energy bunch of trapped electrons present at the end of each simulation. Where there were no discernible quasi-mono-energetic population at higher energies, the charge is left declared as N/A.}
   \label{AsymScan1530J}
\end{table}

\section{Conclusion}

We gave reexamined LWFA in the nonlinear self-guided regime. We have shown that self-guiding continues to occur as the plasma density is lowered and the acceleration length increases in Rayleigh lengths. Optimization of this regime for fixed laser energy was investigated showing that the electron energy can be increased if the pulse length is shortened so long as self-guiding is still achieved. In addition, the effects of modifying the longitudinal profile of the laser for a fixed laser energy was also discussed, as well as the effect of combining these changes with a shortened overall pulse length. Future work may involve using more controlled self-injection methods combined with better tuning the re-phasing mechanism of the spike to reduce the loss in the total accelerated charge as well as considering more complicated laser profiles. Rather than approaching the Lu et al. model as a plug-and-play estimate of trapped particle energies given specific laser parameters, these simulations venture to explore the dimensions of parameters implied by the phenomenological processes presented within it, while simultaneously providing the perspective and vocabulary with which to discuss future studies of these optimization methods.

\section{Acknowledgements}

This work was supported by the US National Science Foundation under NSF Grants No. ACI-1339893 and 1500630, and the US Department of Energy under Grants No. DE-SC0010064 and No. DE-SC0014260. The simulations were performed on the UCLA Hoffman 2 and Dawson 2 Clusters, and the resources of the National Energy Research Scientific Computing Center and the Blue Waters through NSF ACI-1440071.

\bibliographystyle{aipauth4-1}
\bibliography{optimize_laser}

\end{document}